\RequirePackage[hyphens]{url}

\documentclass{article}
\usepackage{ismir,amsmath,cite,url}
\usepackage{graphicx}
\usepackage{color}

\title{A Data-Driven Approach to Smooth Pitch Correction for Singing Voice in Pop Music}

\multauthor
{Sanna Wager\hspace{1cm} Lijiang Guo \hspace{1cm} Aswin Sivaraman \hspace{1cm} Minje Kim}
{
School of Informatics, Computing, and Engineering, Indiana University, USA\\
{\tt\small \{scwager, lijguo, asivara, minje\}@indiana.edu}
}
\def\authorname{Sanna Wager, Lijiang Guo, Aswin Sivaraman, Minje Kim}

\begin{document}

\maketitle

\begin{abstract}
In this paper, we present a machine-learning approach to pitch correction for voice in a karaoke setting, where the vocals and accompaniment are on separate tracks and time-aligned. The network takes as input the time-frequency representation of the two tracks and predicts the amount of pitch-shifting in cents required to make the voice sound in-tune with the accompaniment. It is trained on examples of semi-professional singing. The proposed approach differs from existing real-time pitch correction methods by replacing pitch tracking and mapping to a discrete set of notes---for example, the twelve classes of the equal-tempered scale---with learning a correction that is continuous both in frequency and in time directly from the harmonics of the vocal and accompaniment tracks. A Recurrent Neural Network (RNN) model provides a correction that takes context into account, preserving expressive pitch bending and vibrato. This method can be extended into unsupervised pitch correction of a vocal performance---popularly referred to as autotuning.
\end{abstract}

\section{Introduction}
\label{sec:intro}


Pitch correction of vocals is a commonly desired feature in live vocal performances, notably, karaoke. The task is not straightforward. Vocalists may have different levels of sharpness or flatness (that is, distances above or below the intended pitch, respectively) from note to note, making it difficult to know exactly what pitch they intend to hit. Additionally, singers commonly use pitch bending and vibrato for expressive means: Pitch correction methods should preserve intentional variations. The first commercial pitch detection and correction apparatus was developed and patented in 1997 by the CEO of Antares Audio Technologies, Dr. Harold (Andy) Hildebrand. This technology, trademarked as Auto-Tune, tracks a singer's frequency and adjusts the output audio according to various levels of user input. As described in the user manual \cite{antares:2016} and in recent work on continuous score-coded pitch correction by Salazar \textit{et al.} \cite{salazar2015continuous}, the vocals can be either tuned automatically, in which case each vocal note is pitch-shifted to the nearest note in the user-input set of pitches (scale), or manually, in which case a recording engineer must use the plugin's interface to move each note to the desired pitch or use a score as an input. We focus on the automatic approach. 

The most common scale is the equal-tempered scale, in which each pitch $p$ belongs to the set $[0, 1, ..., 127]$ and its frequency in Hertz is defined as $440*2^{\frac{p-69}{12}}$. Some users prefer a finer resolution and include more than twelve pitches per octave, or use intervals of varying sizes between pitches, as is the case in just temperament. In all cases, pitch is discretized to a small set of values. This model does not directly take into account a singer's intentional pitch variation for expressive means, using techniques such as glissando or vibrato. In order to avoid flattening the singer's pitch to the note, producing a robotic sound, Auto-Tune introduces a "time-lag" parameter that corrects pitch gradually, thus creating a user-adjustable tradeoff between preservation of pitch variation and accuracy. Additionally, real-time user input can add pre-defined pitch variations such as vibrato. In all cases, the pitch is centered around the discrete $p$, which may not always be the most artistically desirable musical choice. For example, empirical studies have shown that musicians compress small musical intervals and stretch large ones, thus assigning different frequencies to the same note depending on the context. Furthermore, soloists often center their frequency at a higher level than the accompaniment, possibly in order to stand out \cite{kantorski1986string} \cite{rakowski1985perception}. More generally, frequency and perceived pitch are often slightly different \cite{parncutt2018psychocultural}. We describe a continuous-scale pitch correction algorithm that can take these musical choices and phenomena of physics and the human auditory system into account. While the original Auto-Tune is most suitable for music following the Western twelve-tone scale, the proposed algorithm additionally adapts easily to other musical-cultural contexts, with different scales or more fluidly varying pitch. 

\begin{figure*}[h]
\centerline{\framebox{
\includegraphics[width=\textwidth]{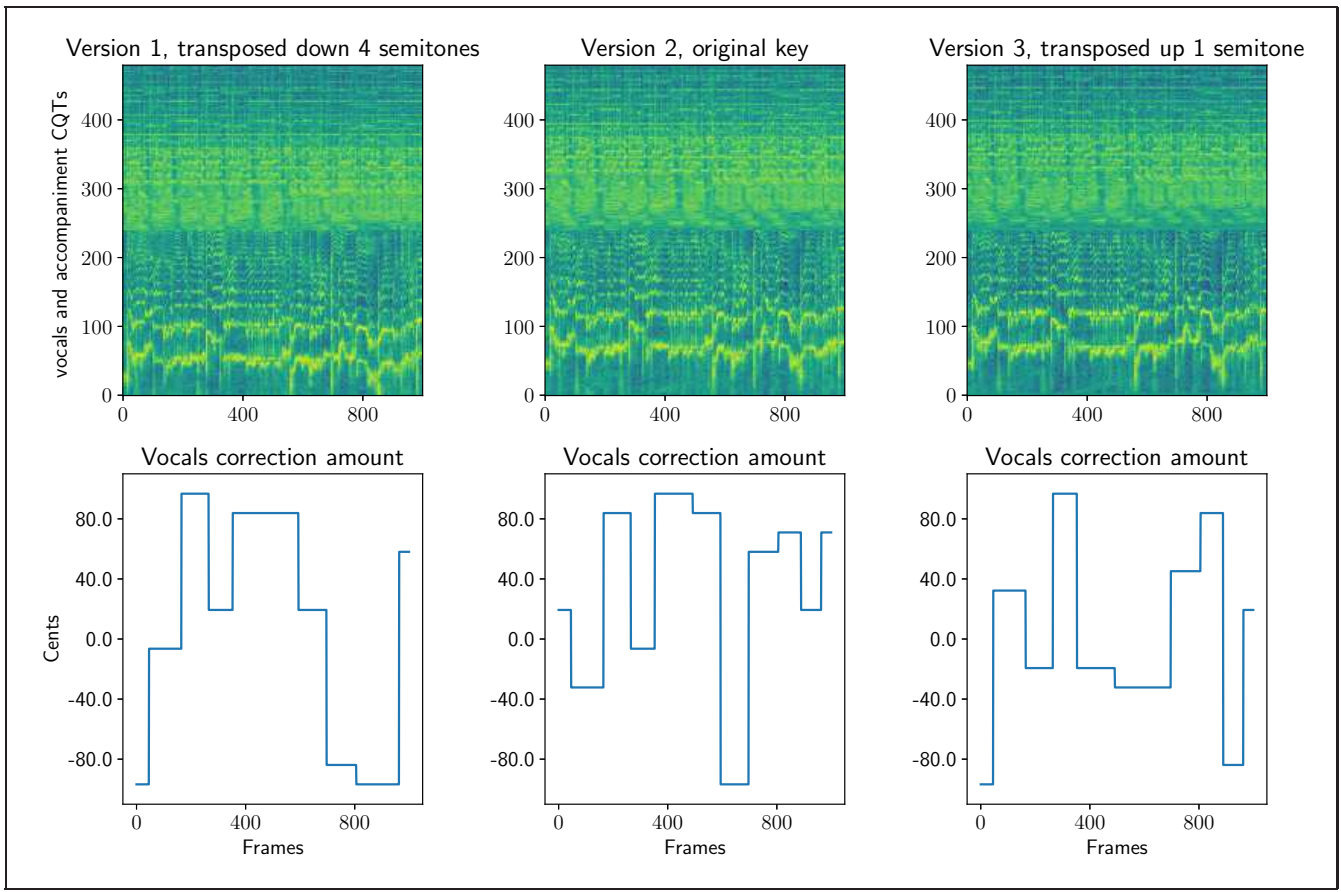}}}
\caption{Input data (top) and corresponding training labels (bottom). The CQT consists of the accompaniment at the top and the vocals at the bottom. The first and third examples were globally shifted in pitch (both the accompaniment and vocals were shifted by the same amount across the whole song) for data augmentation. The correction amount displays by how many cents each segment of the vocals needs to be pitch shifted to return to the original in-tune version.}
\label{fig:input data}
\end{figure*}

The proposed model predicts pitch correction based on a machine-learning approach. Its input consists of vocal and instrumental time-frequency representations and output consists of a sequence of frame-wise pitch corrections that map to the continuous frequency (Hz) scale, avoiding pitch discretization. A Recurrent Neural Network (RNN) allows the model to take the context into account when predicting pitch correction, which is crucial for two reasons. First, audio has a high level of variability and is very noisy due to unpitched sounds (percussion in the accompaniment, consonants in the voice), which calls for a recurrent structure in the model to robustly keep the memory of the notes of interest. Second, a singer's note or melodic contour can last a second or multiple seconds, and the choice of frequency (tuning) at a given frame depends highly on this time interval. Prediction of pitch correction without taking context into account would be difficult or impossible even for a musically trained person. Our model is trained on advanced and professional singing, where artists use pitch deviation, and thus learns to adjust a singer's pitch when it is out of tune without affecting expressive choices. 

RNNs with gating techniques have been successfully adopted in audio signal processing and music information retrieval applications. Long Short-Term Memory networks (LSTMs) are one of the most successful RNN types thanks to their gates and controlled memory cells that have helped resolve the gradient vanishing problem during backpropagation through time \cite{hochreiter1997long}. LSTMs have been widely used in many tasks where modeling audio signals as time series data is important---improvising monophonic melodies given an accompaniment \cite{eck2002finding}, finding structure from music \cite{eck2008learning}, onset detection \cite{eyben2010universal}, source separation \cite{weninger2015speech}, just to name a few. More recently, Gated Recurrent Units (GRUs) have shown comparable performance to LSTMs while providing a simpler structure \cite{chung2014empirical}, and have also been adopted in the Music Information Retrieval (MIR) domain, e.g., for music classification \cite{choi2017convolutional}. 

In this work we choose GRUs with no loss of generality. The proposed model differs from the other RNN models for MIR applications in the sense that it is designed to learn the adequate frequency-wise alignment between the professionally sung main melody and the professionally played accompaniment, where the alignment would not always be obvious if not for the long-term sequence analysis. The main contributions of this paper are as follows:
\begin{itemize}
    \item For the experiment we labeled the SiSEC MUS data set \cite{SiSEC17} with 100 professionally recorded pop songs to identify the parts with monophonic, pitched singing voice, because the ``vocals'' track tends to include chorus sections. We make this annotation public.\footnote{The URL will be available after the review process.}
    \item To our best knowledge, the proposed method is the first data-driven approach to correcting the pitch of singing voice according to the accompaniment. 
    \item To this end, we employ a GRU network to build a baseline system for this new problem. We hope that the adaptive nature of the proposed method can lead to many interesting follow-ups such as a system specialized for a particular main instrument or genre.  
\end{itemize}



\begin{figure*}[h]
\centerline{\framebox{
    \includegraphics[width=0.5\textwidth]{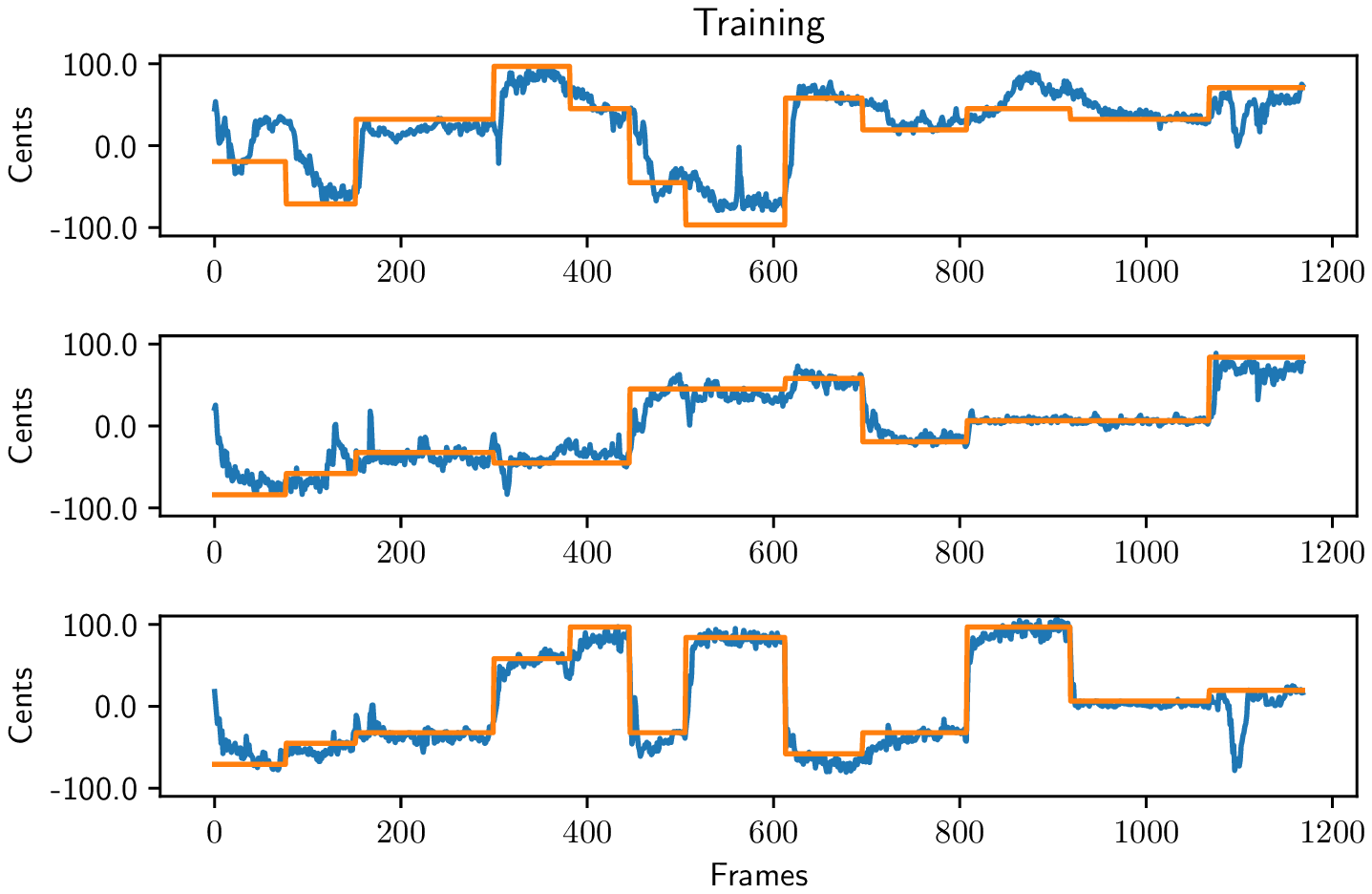}
    \includegraphics[width=0.5\textwidth]{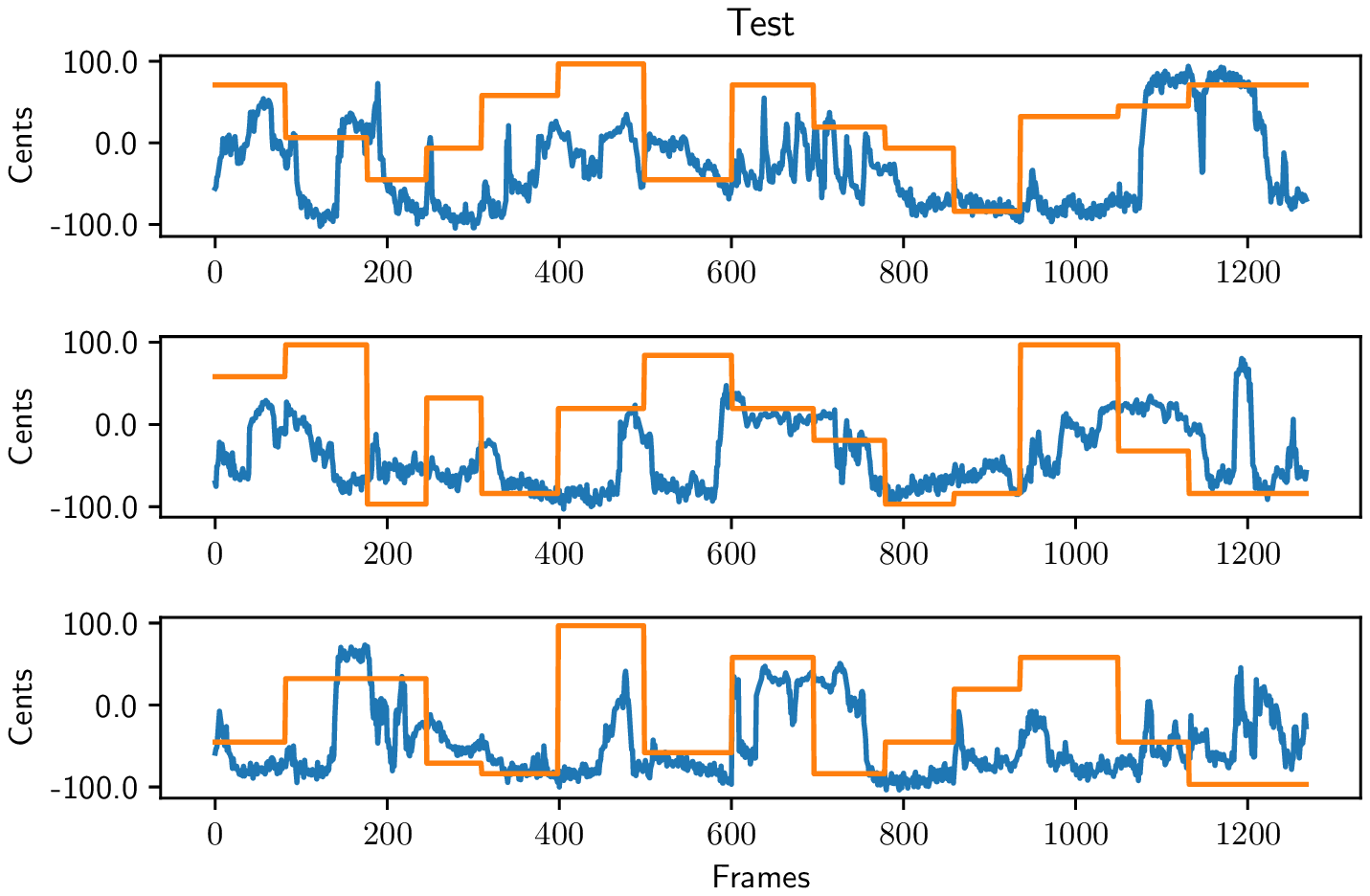}}}
    \caption{Sample predictions on three different randomly pitch shifted versions of a training song and of a test song with a four-layer RNN. The state size was set to 128 for each layer. While the predictions for the training demonstrate some learning, the test results indicate overfitting.}
    \label{fig:results}
\end{figure*}

\section{The Data}
\label{sec:data}
Our program is designed for karaoke data, where the monophonic vocal track is separate from the accompaniment. We used studio recordings of semi-professional singers as examples of in-tune singing, although they are not always perfectly in-tune, and generated multiple de-tuned versions of each song, training the model to predict the amount by which the pitch was shifted. To this end, we used unmixed tracks from the SiSEC MUS dataset \cite{SiSEC17}, which provides four stems of 100 professionally recorded tracks of music of varying styles. We only used songs of musical genres where vocals were melodic, keeping 66 songs. The retained songs still contained multiple genres, vocal and instrumental styles. The ``vocals'' tracks in the SiSEC MUS dataset often contained more than one voice. We retained only monophonic singing, also discarding silent sections. We generated the accompaniment by mixing the ``bass'', ``drums'', and ``other'' tracks. In total, we had approximately 66000 frames, or 25 minutes of music. 

\subsection{De-tuning process}
In order to generate training examples for the model, we need two performances of the same song, where one version of the vocals is out of tune and the other is considered a good performance. The versions should have identical timing and expressive gestures. We choose to generate out-of-tune singing by applying pitch shifts to in-tune singing while keeping the accompaniment fixed. We use the simplifying assumption that a singer has one intended pitch for every note, and that the amount of de-tuning can only change at note boundaries, remaining otherwise constant. We deviate by only up to one semitone (100 cents) to avoid ambiguity of choosing between semitones and larger intervals. The automatic Antares Auto-Tune has a similar scope as it centers the pitch around the nearest note. We make a second simplifying assumption, namely, that the amount of de-tuning between notes is independent. Instead of attempting to synthesize these ideally de-tuned versions, which could introduce additional artifacts, we use a pre-defined set of pitch shifts to de-tune the entire vocal track and then shuffle them randomly at the note level at every iteration. We generate 16 different shifts of the vocals, ranging from 100 cents downwards to 100 cents upwards, evenly spaced in the log-scale. During this procedure we assume that 16 evenly distributed pitch shifts for two semitones are fine enough for the network to be generalized to detect the off-pitch amount in between. Given the large variety of vocal styles, instrumentation, and note combinations in the dataset, we apply data augmentation by applying a global pitch shift of up to four semitones to the accompaniment and vocals combined. Given that the CQT bins are logarithmically spaced, this augmentation is easily performed by shifting the CQT vertically. The shifting is applied using Librosa's resampling and phase vocoder utilities \cite{mcfee2015librosa}. Figure \ref{fig:input data} illustrates the structure of the data.

\subsection{Format}
We compute the Constant-Q Transform (CQT) of the vocals and the accompaniment with a resolution of eight bins per note, up to approximately 4khz. For the voice, we use a minimum frequency of 75 Hertz (Hz). The vocals and accompaniment are stacked for a resulting frame dimension of 536. We use non-overlapping frames of 1024 samples with a Hanning window, so one second of audio contains approximately 43 frames. The pitch shifting amount for every frame is recorded as a scalar.

\section{The Model}
\label{sec:model}
Our model is a GRU. We tested for one to four hidden layers with state sizes ranging from 32 to 512 and a sequence length ranging from 43 to 215 frames, equivalent to one to five seconds of audio. We advanced by 20 frames for each batch, thus creating overlap, and used only the last 20 frames of each batch in the final prediction. The learning rate was set to 1e-4 for the Adam optimizer \cite{kingma2014adam} with annealing and early stopping. The batch size was set to 32 sequences, each a different randomly pitch-shifted version of the same song. The activation functions were the hyperbolic tangent (tanh) for the RNN layers and logistic functions for gating. For the output, we used a linear activation function. This avoids having a steep gradient at zero, which would prevent in-tune songs from getting predictions near zero. Our logarithmic-scale shifts ranging from $-100$ to $100$ cents were mapped to the linear scale in the range of $-1$ to $1$. The error function was the average Mean-Squared Error (MSE) between the pitch shift estimate and ground truth over the full sequence. We leave re-synthesis of a pitch-corrected song and use of Signal-To-Noise Ratio (SNR) and other audio quality metrics to future work. The model was built in TensorFlow \cite{tensorflow}.

\section{Experiment Results}
\label{sec:results}
We trained on 61 songs and tested on the remaining 5. Figure \ref{fig:results} displays sample results. The model is able to generalize to different permutations of training data when the model has two or more layers. However, the test loss does not decrease over time. Unknown songs---with a different voice, tonality, accompaniment and instrumentation---are out of the scope of this shallow network's ability to learn frequency relationships across spectrograms. Given the small size of the dataset, a first step is to build a larger dataset that covers more note combinations and/or is more restricted in terms of genre. Additionally, both our representation of a note’s pitch structure and the RNN structure can be further developed.

\section{Conclusion and future work}
This experiment is the first iteration of a deep-learning model that estimates pitch correction for an out-of-tune monophonic input vocal track using an instrumental accompaniment track as reference. Our results on RNNs of two, three, and four layers indicate that the spectral information of an accompaniment and of a vocal track is useful for determining the amount of pitch-correction required at each frame. This project is an initial prototype that we plan to develop into a model robust to variance in voice types, tonality, noise, and accompaniment instrumentation, all of which make challenging the task of learning a relationship between harmonics and detecting an offset. While the current model outputs the amount by which singing should be shifted in pitch, the model can be extended to perform autotuning, either by post-processing the voice recording or by developing the model to directly output the modified audio.

\bibliography{ISMIRtemplate.bib}

\begin{thebibliography}{10}

\bibitem{tensorflow}
M.~Abadi, P.~Barham, J.~Chen, Z.~Chen, A.~Davis, J.~Dean, M.~Devin,
  S.~Ghemawat, G.~Irving, M.~Isard, et~al.
\newblock Tensorflow: A system for large-scale machine learning.
\newblock In {\em OSDI}, volume~16, pages 265--283, 2016.

\bibitem{choi2017convolutional}
K.~Choi, G.~Fazekas, M.~Sandler, and K.~Cho.
\newblock Convolutional recurrent neural networks for music classification.
\newblock In {\em Acoustics, Speech and Signal Processing (ICASSP), 2017 IEEE
  International Conference on}, pages 2392--2396. IEEE, 2017.

\bibitem{chung2014empirical}
J.~Chung, C.~Gulcehre, K.~Cho, and Y.~Bengio.
\newblock Empirical evaluation of gated recurrent neural networks on sequence
  modeling.
\newblock {\em arXiv preprint arXiv:1412.3555}, 2014.

\bibitem{eck2008learning}
D.~Eck and J.~Lapalme.
\newblock Learning musical structure directly from sequences of music.
\newblock {\em University of Montreal, Department of Computer Science, CP},
  6128, 2008.

\bibitem{eck2002finding}
D.~Eck and J.~Schmidhuber.
\newblock Finding temporal structure in music: Blues improvisation with {LSTM}
  recurrent networks.
\newblock In {\em Proc. of the 2002 12th IEEE Workshop on Neural Networks for
  Signal Processing}, pages 747--756. IEEE, 2002.

\bibitem{eyben2010universal}
F.~Eyben, S.~B{\"o}ck, B.~Schuller, and A.~Graves.
\newblock Universal onset detection with bidirectional long-short term memory
  neural networks.
\newblock In {\em Proc. 11th Intern. Soc. for Music Information Retrieval
  Conference, ISMIR, Utrecht, The Netherlands}, pages 589--594, 2010.

\bibitem{hochreiter1997long}
S.~Hochreiter and J.~Schmidhuber.
\newblock Long short-term memory.
\newblock {\em Neural computation}, 9(8):1735--1780, 1997.

\bibitem{kantorski1986string}
V.~J. Kantorski.
\newblock String instrument intonation in upper and lower registers: The
  effects of accompaniment.
\newblock {\em Journal of Research in Music Education}, 34(3):200--210, 1986.

\bibitem{kingma2014adam}
D.~P. Kingma and J.~Ba.
\newblock Adam: A method for stochastic optimization.
\newblock {\em arXiv preprint arXiv:1412.6980}, 2014.

\bibitem{SiSEC17}
A.~Liutkus, F.~St{\"o}ter, Z.~Rafii, D.~Kitamura, B.~Rivet, N.~Ito, N.~Ono, and
  J.~Fontecave.
\newblock The 2016 signal separation evaluation campaign.
\newblock In {\em Latent Variable Analysis and Signal Separation: 13th
  International Conference, LVA/ICA 2017, Grenoble, France}, pages 323--332.
  Springer International Publishing, 2017.

\bibitem{mcfee2015librosa}
B.~McFee, C.~Raffel, D.~Liang, D.~P.~W. Ellis, M.~McVicar, E.~Battenberg, and
  O.~Nieto.
\newblock librosa: Audio and music signal analysis in python.
\newblock In {\em Proc. of the 14th Python in Science Conference}, 2015.

\bibitem{parncutt2018psychocultural}
R.~Parncutt and G.~Hair.
\newblock A psychocultural theory of musical interval: Bye bye {P}ythagoras.
\newblock {\em Music Perception: An Interdisciplinary Journal}, 35(4):475--501,
  2018.

\bibitem{rakowski1985perception}
A.~Rakowski.
\newblock The perception of musical intervals by music students.
\newblock {\em Bulletin of the Council for Research in Music Education}, pages
  175--186, 1985.

\bibitem{salazar2015continuous}
S.~Salazar, R.~A. Fiebrink, G.~Wang, M.~Ljungstr{\"o}m, J.~C. Smith, and P.~R.
  Cook.
\newblock Continuous score-coded pitch correction, September~29 2015.
\newblock US Patent 9,147,385.

\bibitem{weninger2015speech}
F.~Weninger, H.~Erdogan, S.~Watanabe, E.~Vincent, J.~Le~Roux, J.~R. Hershey,
  and B.~Schuller.
\newblock Speech enhancement with {LSTM} recurrent neural networks and its
  application to noise-robust {ASR}.
\newblock In {\em International Conference on Latent Variable Analysis and
  Signal Separation}, pages 91--99. Springer, 2015.

\end{thebibliography}

\end{document}